\documentclass[12pt,a4paper]{article}
\usepackage[left=2.5cm,right=2.5cm,top=3.5cm,bottom=3.5cm,a4paper]{geometry}
\pdfoutput=1 

\usepackage[T1]{fontenc} 
\usepackage[utf8]{inputenc}
\usepackage{amsmath,amssymb,xspace,hyperref,graphicx,color,ifpdf,ifthen,mciteplus}
\usepackage{booktabs,multirow,cite,braket}
\usepackage{bm} 
\usepackage{tikz}
\usetikzlibrary{arrows}
\graphicspath{{HF_revisited_figs/}}

\newboolean{pdflatex}
\setboolean{pdflatex}{true} 

\newboolean{articletitles}
\setboolean{articletitles}{true} 

\newboolean{uprightparticles}
\setboolean{uprightparticles}{false} 

\newboolean{inbibliography}
\setboolean{inbibliography}{false} 

\def\Lc      {{\ensuremath{\Lz^+_\cquark}}\xspace}

\newcommand{\tevc}{\ensuremath{{\mathrm{\,Te\kern -0.1em V\!/}c}}\xspace}
\newcommand{\tevtevcccc}{\ensuremath{{\mathrm{\,Te\kern -0.1em V^2\!/}c^4}}\xspace}
\newcommand{\gevgevcc}{\ensuremath{{\mathrm{\,Ge\kern -0.1em V^2\!/}c^2}}\xspace}
\newcommand{\tevtevcc}{\ensuremath{{\mathrm{\,Te\kern -0.1em V^2\!/}c^2}}\xspace}

\def\s0z   {{\ensuremath{ s_{0,z} }}\xspace}

\newcommand{\Lcpkpi}{\ensuremath{\Lc\to pK^-\pi^+}\xspace}

\newcommand{\mqpk}{\ensuremath{m^2_{pK^-}}\xspace}
\newcommand{\mqkpi}{\ensuremath{m^2_{K^-\pi^+}}\xspace}
\newcommand{\mqppi}{\ensuremath{m^2_{p\pi^+}}\xspace}

\usepackage{xspace} 
\usepackage{caption} 

\usepackage{graphicx}  
\usepackage{color}
\usepackage{colortbl}

\usepackage{amsmath} 
\usepackage{amssymb}
\usepackage{amsfonts}
\usepackage{upgreek} 

\usepackage{xspace} 
\usepackage{upgreek}

\def\MagUp {\mbox{\em Mag\kern -0.05em Up}\xspace}

\ifthenelse{\boolean{uprightparticles}}%
{

 \def\PDelta      {\ensuremath{\Delta}\xspace}                 
 \def\PXi      {\ensuremath{\Xi}\xspace}                 
 \def\PLambda      {\ensuremath{\Lambda}\xspace}                 
 \def\PSigma      {\ensuremath{\Sigma}\xspace}                 
 \def\POmega      {\ensuremath{\Omega}\xspace}                 
 \def\PUpsilon      {\ensuremath{\Upsilon}\xspace}                 
 
 \def\PB      {\ensuremath{\mathrm{B}}\xspace}                 
                  
 \def\PD      {\ensuremath{\mathrm{D}}\xspace}

 \def\PK      {\ensuremath{\mathrm{K}}\xspace}

 \def\Pc      {\ensuremath{\mathrm{c}}\xspace}

 \def\Pi      {\ensuremath{\mathrm{i}}\xspace}

}
{

 \mathchardef\PDelta="7101
 \mathchardef\PXi="7104
 \mathchardef\PLambda="7103
 \mathchardef\PSigma="7106
 \mathchardef\POmega="710A
 \mathchardef\PUpsilon="7107
                  
 \def\PB      {\ensuremath{B}\xspace}                 
                  
 \def\PD      {\ensuremath{D}\xspace}

 \def\PK      {\ensuremath{K}\xspace}

 \def\Pc      {\ensuremath{c}\xspace}

 \def\Pi      {\ensuremath{i}\xspace}

}

\def\cquark    {{\ensuremath{\Pc}}\xspace}

  \def\Kbar    {{\kern 0.2em\overline{\kern -0.2em \PK}{}}\xspace}

\def\KorKbar    {\kern 0.18em\optbar{\kern -0.18em K}{}\xspace}

  \def\Dbar    {{\kern 0.2em\overline{\kern -0.2em \PD}{}}\xspace}

\def\DorDbar    {\kern 0.18em\optbar{\kern -0.18em D}{}\xspace}

\def\Bbar    {{\ensuremath{\kern 0.18em\overline{\kern -0.18em \PB}{}}}\xspace}

\def\BorBbar    {\kern 0.18em\optbar{\kern -0.18em B}{}\xspace}

  \def\Y#1S{\ensuremath{\PUpsilon{(#1S)}}\xspace}

\def\Deltares    {{\ensuremath{\PDelta}}\xspace}

\def\Lz          {{\ensuremath{\PLambda}}\xspace}
\def\Lbar        {{\ensuremath{\kern 0.1em\overline{\kern -0.1em\PLambda}}}\xspace}
\def\LorLbar    {\kern 0.18em\optbar{\kern -0.18em \PLambda}{}\xspace}

\def\Lc      {{\ensuremath{\Lz^+_\cquark}}\xspace}

\def\to                 {\ensuremath{\rightarrow}\xspace}

\def\C#1      {\ensuremath{\mathcal{C}_{#1}}\xspace}                       
\def\Cp#1     {\ensuremath{\mathcal{C}_{#1}^{'}}\xspace}                    
\def\Ceff#1   {\ensuremath{\mathcal{C}_{#1}^{\mathrm{(eff)}}}\xspace}        
\def\Cpeff#1  {\ensuremath{\mathcal{C}_{#1}^{'\mathrm{(eff)}}}\xspace}       
\def\Ope#1    {\ensuremath{\mathcal{O}_{#1}}\xspace}                       
\def\Opep#1   {\ensuremath{\mathcal{O}_{#1}^{'}}\xspace}                    

\newcommand{\tev}{\ensuremath{\mathrm{\,Te\kern -0.1em V}}\xspace}
\newcommand{\gev}{\ensuremath{\mathrm{\,Ge\kern -0.1em V}}\xspace}
\newcommand{\mev}{\ensuremath{\mathrm{\,Me\kern -0.1em V}}\xspace}
\newcommand{\kev}{\ensuremath{\mathrm{\,ke\kern -0.1em V}}\xspace}
\newcommand{\ev}{\ensuremath{\mathrm{\,e\kern -0.1em V}}\xspace}
\newcommand{\gevc}{\ensuremath{{\mathrm{\,Ge\kern -0.1em V\!/}c}}\xspace}
\newcommand{\mevc}{\ensuremath{{\mathrm{\,Me\kern -0.1em V\!/}c}}\xspace}
\newcommand{\gevcc}{\ensuremath{{\mathrm{\,Ge\kern -0.1em V\!/}c^2}}\xspace}
\newcommand{\gevgevcccc}{\ensuremath{{\mathrm{\,Ge\kern -0.1em V^2\!/}c^4}}\xspace}
\newcommand{\mevcc}{\ensuremath{{\mathrm{\,Me\kern -0.1em V\!/}c^2}}\xspace}

\def\gsim{{~\raise.15em\hbox{$>$}\kern-.85em
          \lower.35em\hbox{$\sim$}~}\xspace}
\def\lsim{{~\raise.15em\hbox{$<$}\kern-.85em
          \lower.35em\hbox{$\sim$}~}\xspace}

\def\tell1  {TELL1\xspace}
\def\ukl1   {UKL1\xspace}

\newcommand{\ie}{\mbox{\itshape i.e.}\xspace}

\usepackage{cite} 
\usepackage{mciteplus}

\begin{document}

\title{\boldmath \textbf{Helicity amplitudes for generic multi-body particle decays featuring multiple decay chains}}

 \author{\textbf{Daniele Marangotto}\footnote{E-mail: \texttt{daniele.marangotto@unimi.it}}\\[2ex]Universit\`a degli Studi di Milano and INFN Milano,\\[2ex]Via Celoria 16, 20133 Milano, Italy}
 
\maketitle

\begin{abstract}
We present the general expression of helicity amplitudes for generic multi-body particle decays characterised by multiple decay chains. This is achieved by addressing for the first time the issue of the matching of final particle spin states among different decay chains in full generality for generic multi-body decays, proposing a method able to match the exact definition of spin states relative to the decaying particle ones. We stress the importance of our result by showing that one of the matching method used in the literature is incorrect, leading to amplitude models violating rotational invariance. The results presented are therefore relevant for performing numerous amplitude analysis, notably those searching for exotic structures like pentaquarks.

\end{abstract}

\section{Introduction}
\label{sec:intro}

The helicity formalism, proposed by Jacob and Wick~\cite{JacobWick} in 1959 to treat relativistic processes involving particles with spin, is still one of the most important tools for performing amplitude analyses of particle decays. To date, complex amplitude analyses involving final-state particles with spin and multiple decay chains have been performed, especially for the search of new resonant structures. Pentaquark searches are a typical example: a pentaquark involves at least one baryon in the final state and introduces an additional decay chain. For instance, pentaquark states were discovered by the LHCb collaboration performing an amplitude analysis of the $\Lambda^0_b \to J/\psi pK^-$ baryon decay~\cite{LHCb-PAPER-2015-029}.

However, a consistent definition of final particle spin states for these kind of decays turned out to be an issue, since the definition of helicity states is different for different decay chains. Various solutions to match final particle spin states have been proposed~\cite{Mizuk:2008me,LHCb-PAPER-2015-029,Chen:2017gtx,Mikhasenko:2019rjf}, but none addressed the problem in full generality for generic multi-body decays. In this paper we present a general method for matching spin states, obtained requiring that, for any decay chain, final particle states are defined by the same Lorentz transformations relatively to the decaying particle spin states.

To this end, we first review the definition of spin states in quantum mechanics in Sect.~\ref{sec:phase_definition}, with a particular attention to their phase specifications. The key point we want to stress is that the relative phases among sets of spin states linked by rotations are fully specified by the transformations applied. Therefore, phase differences like those arising when spin states are rotated with respect to their quantisation axis, or like the change of sign under $2\pi$ angle rotations of fermion states can not be neglected in helicity amplitudes.

Then, we revisit the helicity formalism as originally proposed by Jacob and Wick~\cite{JacobWick}, highlighting the different treatment of daughter particle helicity states in two-body processes. We also propose a simpler definition of two-particle helicity states than the standard one, which allows for an easier matching of final particle spin states.

In Sect.~\ref{sec:helicity_amplitudes} we present how to write helicity amplitudes with a consistent definition of final particle spin states for different decay chains, applicable to any multi-body decay topology. We explicitly derive helicity amplitudes for three-body decays.

We stress the need for a consistent definition of final particle spin states in Sect.~\ref{sec:effects_phase}. First, we discuss the consequences of an incorrect phase introduced between amplitudes describing different decay chains on the decay distributions, showing they produce observable effects on the decay distributions via interference terms. Next, we perform a numerical study on \Lcpkpi helicity amplitude models featuring different methods to match final particle spin states, checking a general property of the decay distributions following from rotational invariance. We show how the method employed for the amplitude analyses Refs.~\cite{Mizuk:2008me,LHCb-PAPER-2015-029} is incorrect, leading to amplitude models violating rotational invariance, while that proposed in this article fully satisfies rotational symmetry.


\section{On spin states definition}
\label{sec:phase_definition}
In this section, we review the definition of spin states in quantum mechanics underlining the importance of their phase specification, which will be needed for the upcoming discussion of multi-body particle decays in the helicity formalism.

In quantum mechanics, the spin of a particle is described by a vector of spin operators $\bm{\hat{S}}=(\hat{S}_x,\hat{S}_y,\hat{S}_z)$, which defines a right-handed spin coordinate system $(x,y,z)$. The spin states $\Ket{s,m}$ are defined as the simultaneous eigenstates of the spin squared modulus $\hat{S}^2$ and $\hat{S}_z$, with eigenvalues $s(s+1)$ and $m$, respectively. The $z$ axis is called quantisation axis, while $x$ and $y$ axes will be named orthogonal axes.

The choice of the orthogonal axes specifies the relative phases among spin states, which can be conventionally chosen by defining the action of the ``ladder'' operators $\hat{S}_{\pm} = \hat{S}_x \pm i \hat{S}_y$ transforming $\ket{s,m}$ into $\ket{s,m\pm 1}$ eigenstates, see \textit{e.g.} Ref.~\cite{Sakurai}. The overall phase of the spin states is undefined and can be chosen arbitrarily.

Now, let's consider a set of spin states $\ket{s,m}'$ defined relative to the original one $\ket{s,m}$ by applying a rotation $\hat{R}$,
\begin{equation}
\ket{s,m}' = \hat{R} \ket{s,m}.
\end{equation}
The rotation $R$ defines the relative phases among the two sets: for instance, a rotation around the $z$ axis of angle $\alpha$ introduces a phase difference between original and rotated spin states,
\begin{align}
\ket{s,m}' &= \hat{R}_{z}(\alpha) \ket{s,m} = e^{-i\alpha \hat{S}_{z}} \ket{s,m} = e^{-i\alpha m} \ket{s,m}.
\label{eq:state_transformation}
\end{align}
Therefore, once an overall phase for the original set of spin states is conventionally chosen, that of the rotated spin states is defined by the rotation. In other words, the rotated states are completely defined in terms of the original states and the rotation.

The fact that the expectation values of the spin operators transform as a vector under rotations, see \textit{e.g.} Ref.~\cite{Sakurai}, can give the deceptive impression that one can represent rotations applied to spin states in the usual Cartesian space, while spin states transform under spin $s$ representations of the $SU(2)$ group\footnote{This is why graphical descriptions are not used in the present article, though they are widely used in the literature.}. 
For instance, it is well known that fermion states change sign for a $2\pi$ angle rotation around any axis $i$,
\begin{equation}
\hat{R}_{i}(2\pi) \ket{s,m} = e^{-2i\pi \hat{S}_{i}} \ket{s,m} = (-1)^{2s} \ket{s,m},
\label{eq:2pi_rotation_fermion}
\end{equation}
even if the spin operator expectation values do not change.

When considering sets of spin states relatively defined by rotations it is important to take into account their relative phase differences, since interference effects can make them observable quantities. In this article, we will show how to properly consider the spin state definition in the case of multi-body particle decays with different intermediate states in the helicity formalism, Sect.~\ref{sec:helicity_amplitudes}, and the consequences that an incorrect treatment of spin state definitions have on particle decay distributions, Sect.~\ref{sec:effects_phase}.

\section{Helicity formalism revisited}
\label{sec:helicity_formalism_rev}
In this section we revisit the helicity formalism~\cite{JacobWick}, developed to overcome problems related to the treatment of spin in relativistic processes. In particular, we highlight the different role played by the daughter particles in two-body processes and the importance of consistently specifying the definition of their helicity states, including phases. This is an aspect almost neglected so far, which becomes essential for a correct treatment of decays characterised by multiple interfering decay chains. For a clearer treatment of such aspects and a simpler matching of the final particle spin definitions among different decay chain, Sect.~\ref{sec:helicity_amplitudes}, we also propose a different way to express two-particle helicity states, which ease the control of their definitions. A review of the description of relativistic processes involving particles with spin is reported in Appendix~\ref{sec:spin_relativistic}; there, the definition of canonical and helicity states used throughout the article are presented.

The key point underlying the helicity formalism is the invariance of helicity under rotations, exploited to construct two-particle states which are eigenstates of total angular momentum. Indeed, under rotations, both spin states and the momentum expressing their quantisation axis rotate, so that the projection of the particle spin on the momentum is unchanged.

Let's consider a two-body decay $A\to 1,2$. The particle 1 helicity states $\Ket{\bm{p}^A_1,s_1,\lambda_1}$ are defined in the helicity system, see Eq.~\eqref{eq:helicity_state_definition},
\begin{equation}
S^H_1 = L(-p^A_1 \bm{z})R(0,-\theta_1,-\phi_1) S_A,
\label{eq:particle_1_helicity_state}
\end{equation}
with $\bm{p}^A_1$ the particle 1 momentum in the $A$ spin reference rest frame $S_A$ and $\theta_1,\phi_1$ its spherical coordinates.

The particle 2 helicity states $\Ket{\bm{p}^A_2,s_2, \lambda_2}$ are defined in the helicity system
\begin{align}
S^H_2 &= L(-p^A_2 \bm{z})R(0,-\theta_2,-\phi_2) S_A,
\label{eq:particle_2_helicity_state}
\end{align}
which is reached by a different rotation with respect particle 1 helicity states, so that the direct product of particle 1 and 2 states can not be related to the total angular momentum $S_A$. Exploiting $\bm{p}^A_1=-\bm{p}^A_2$, in Eqs.~(13),~(14) of Ref.~\cite{JacobWick} the direct product of daughter particles helicity states is defined as
\begin{align}
\Ket{p^A_1,\theta_1,\phi_1, \lambda_1,\lambda_2} \equiv \Ket{\bm{p}^A_1,s_1, \lambda_1}
\otimes (-1)^{s_2-\lambda_2} \exp(-i\pi \hat{J}_y) \Ket{\bm{p}^A_2,s_2, \lambda_2}.
\label{eq:plane_wave_helicity}
\end{align}
These states can be now related to two-particle states with definite value of total angular momentum, denoted $\Ket{p^A_1,J,M,\lambda_1,\lambda_2}$, as
(here with the $\psi=0$ convention described in Appendix~\ref{sec:spin_relativistic}),
\begin{align}
\Ket{p^A_1,\theta_1,\phi_1, \lambda_1,\lambda_2} &= \sum_{J,M} \sqrt{\frac{2J+1}{4\pi}} D^{J}_{M,\lambda_1-\lambda_2}(\phi_1,\theta_1,0) \Ket{p^A_1,J,M,\lambda_1,\lambda_2}.
\label{eq:plane_spherical_wave_relation}
\end{align}
This expression allows to write the $A\to 1,2$ decay amplitude as
\begin{align}
\mathcal{A}_{m_A,\lambda_1,\lambda_2}(\theta_1,\phi_1) &= \Braket{p^A_1,\theta_1,\phi_1, \lambda_1,\lambda_2|\hat{T}|s_A,m_A} \nonumber\\
&= \mathcal{H}_{\lambda_1,\lambda_2} D^{*s_A}_{m_A,\lambda_1-\lambda_2}(\phi_1,\theta_1,0).
\label{eq:two_body_amplitude}
\end{align}
in which $\Ket{s_A,m_A}$ are the $A$ spin states defined in the $S_A$ system, $\hat{T}$ is the transition operator and
\begin{equation}
\mathcal{H}_{\lambda_1,\lambda_2} \equiv \Braket{J=s_A,M=m_A,\lambda_1,\lambda_2|\hat{T}|s_A,m_A},
\label{eq:helicity_couplings}
\end{equation}
are complex numbers called helicity couplings, describing the decay dynamics.
Note that the key point underlying Eq.~\eqref{eq:plane_spherical_wave_relation} is that the state $\Ket{p^A_1,0,0, \lambda_1,\lambda_2}$ (with $\bm{p}^A_1$ aligned with the $z$ axis) is an eigenstate of $\hat{J}_z$ with eigenvalue $\lambda_1-\lambda_2$, which is then rotated to 
\begin{equation}
\Ket{p^A_1,\theta_1,\phi_1, \lambda_1,\lambda_2} = \hat{R}(\phi_1,\theta_1,0)\Ket{p^A_1,0,0, \lambda_1,\lambda_2}.
\end{equation}

We note en passant that, using the properties of Wigner $D$-matrices Eq.~\eqref{eq:D_matrix_properties}, the two-body amplitude can be rewritten as
\begin{equation}
\mathcal{A}_{m_A,\lambda_1,\lambda_2}(\theta_1,\phi_1) = \mathcal{H}_{\lambda_1,\lambda_2} D^{s_A}_{\lambda_1-\lambda_2,m_A}(0,-\theta_1,-\phi_1),
\label{eq:two_body_amplitude_rewritten}
\end{equation}
so that, comparing with Eq.~\eqref{eq:euler_rotation_spin_states}, the Wigner $D$-matrix is indeed the representation of the helicity rotation $R(0,-\theta_1,-\phi_1)$ aligning the $\bm{p}^A_1$ momentum to the $z$ axis on the $A$ particle spin states $\Ket{s_A,m_A}$.

Let's now consider the particle 2 state in Eq.~\eqref{eq:plane_wave_helicity}: the rotation $\exp(-i\pi \hat{S}_y)$ acts on the helicity state inverting the $z$ axis direction; therefore the particle 2 states entering the amplitude Eq.~\eqref{eq:plane_spherical_wave_relation} are actually opposite-helicity states, which represent spin projection eigenstates in the direction opposite to $\bm{p}^A_1$. This different role of particle 1 and 2 states must be properly considered when these particles have a subsequent decay: the amplitude for the particle 2 decay must take into account that it is not referred to $\Ket{\bm{p}^A_2,s_2, \lambda_2}$ states but to those obtained applying the inversion $\exp(-i\pi \hat{S}_y)$ and the phase factor $(-1)^{s_2-\lambda_2}$. 

We stress that these tricky aspects related to the helicity formalism have been neglected or underestimated so far, because for simple processes (like decays via single decay chains or involving spinless particles) they do not have consequences on the decay distributions. However, they matter for the treatment of the more general decays considered in this article.

To take into account in a cleaner way the different role of particle 1 and 2 in the helicity formalism we propose a simpler definition of the two-particle state Eq.~\eqref{eq:plane_wave_helicity}, which allows for an easier matching of final particle spin definitions among different decay chain, Sect.~\ref{sec:helicity_amplitudes}.
We define the two-particle product state as
\begin{equation}
\Ket{p^A_1,\theta_1,\phi_1, \lambda_1,\bar{\lambda}_2} = \Ket{\bm{p}^A_1,s_1, \lambda_1} \otimes \Ket{\bm{p}^A_2,s_2, \bar{\lambda}_2},
\label{eq:plane_wave_helicity_corrected}
\end{equation}
in which $\ket{\bm{p}^A_2,s_2,\bar{\lambda}_2}$ represent spin projection eigenstates in the direction opposite to $\bm{p}^A_2$. The operator $\hat{\bar{\lambda}}$ is the opposite of the helicity,
\begin{equation}
\hat{\bar{\lambda}} = -\hat{\bm{S}}\cdot\frac{\bm{p}}{p},
\end{equation}
and the opposite-helicity reference system of particle 2 is defined by
\begin{equation}
S^{OH}_2 = L(p^A_2 \bm{z})R(0,-\theta_1,-\phi_1) S_A,
\label{eq:opposite_helicity_state_definition}
\end{equation}
that is, the particle 2 rest frame is reached by boosting along its momentum $\bm{p}^A_2$ pointing in the direction opposite to the $z$ axis. Comparing Eqs.~\eqref{eq:particle_1_helicity_state} and~\eqref{eq:opposite_helicity_state_definition}, we see both particle 1 and particle 2 states are obtained from the same rotation $R(0,-\theta_1,-\phi_1)$, so that their spin is referred to the ``same'' spin reference system (they only differ by a boost along the $z$ axis), including the same definition of the orthogonal axes.

It is therefore possible to define eigenstates of total angular momentum $\Ket{p^A_1,J,M,\lambda_1,\bar{\lambda}_2}$ similarly as before, and Eq.~\eqref{eq:plane_spherical_wave_relation} holds with the substitution $-\lambda_2 \to \bar{\lambda}_2$,
\begin{align}
\Ket{p^A_1,\theta_1,\phi_1, \lambda_1,\bar{\lambda}_2} &= \sum_{J,M} \sqrt{\frac{2J+1}{4\pi}} D^J_{M,\lambda_1+\bar{\lambda}_2}(\phi_1,\theta_1,0) \Ket{p^A_1,J,M,\lambda_1,\bar{\lambda}_2}.
\label{eq:plane_spherical_wave_relation_corrected}
\end{align}
The two-body decay amplitude becomes
\begin{align}
\mathcal{A}_{m_A,\lambda_1,\bar{\lambda}_2}(\theta_1,\phi_1) &= \Braket{p^A_1,\theta_1,\phi_1, \lambda_1,\bar{\lambda}_2|\hat{T}|s_A,m_A} \nonumber\\
&= \mathcal{H}_{\lambda_1,\bar{\lambda}_2} D^{*s_A}_{m_A,\lambda_1+\bar{\lambda}_2}(\phi_1,\theta_1,0),
\label{eq:two_body_amplitude_corrected}
\end{align}
and the helicity values allowed by angular momentum conservation are
\begin{equation}
|\lambda_1|\leq s_1,\hspace{1cm} |\bar{\lambda}_2|\leq s_2,\hspace{1cm} |\lambda_1+\bar{\lambda}_2|\leq s_A.
\label{eq:allowed_helicity_couplings}
\end{equation}

The amplitudes Eqs.~\eqref{eq:two_body_amplitude} and \eqref{eq:two_body_amplitude_corrected} are the same but for the substitution $\lambda_2\leftrightarrow\bar{\lambda}_2$, so why bother with a new state definition? The difference is in the definition of particle 2 states: in the standard formulation Eq.~\eqref{eq:plane_wave_helicity} the particle 2 opposite-helicity state is obtained inverting the helicity one, applying two rotations to the initial system $S_A$ plus a phase; in our definition it is just defined by a single rotation from $S_A$. Our choice simplifies both the writing of particle 2 subsequent decay amplitudes and the matching of final particle spin states among different decay chains.

For the purpose of Sect.~\ref{sec:helicity_amplitudes}, it is useful to derive the relation between opposite-helicity and canonical states, the analogue of Eq.~\eqref{eq:canonical_helicity_relation} for helicity states. It is obtained applying Eq.~\eqref{eq:boost_decomposition} along with the relations $\bm{p}^A_2 = -\bm{p}^A_1$, $p^A_2 = p^A_1$, to the definition of canonical states Eq.~\eqref{eq:canoncial_state_definition},
\begin{align}
S^C_2 &= L(-\bm{p}^A_2) S_A \nonumber\\
&= L(\bm{p}^A_1) S_A \nonumber\\
&= R(\phi_1,\theta_1,0) L(p^A_1 \bm{Z}) R(0,-\theta_1,-\phi_1) S_A\nonumber\\
&= R(\phi_1,\theta_1,0) L(p^A_2 \bm{Z}) R(0,-\theta_1,-\phi_1) S_A\nonumber\\
&= R(\phi_1,\theta_1,0) S^{OH}_2.
\label{eq:canonical_opposite_helicity_relation}
\end{align}
The rotation is indeed the same as the one from the helicity to the canonical system of particle 1, see Eq.~\eqref{eq:canonical_helicity_relation},
\begin{equation}
S^C_1 = R(\phi_1,\theta_1,0) S^{H}_1.
\end{equation}

\section{Helicity amplitudes for generic multi-body particle decays featuring multiple decay chains}
\label{sec:helicity_amplitudes}
In this section we present how helicity amplitudes for generic multi-body particle decays characterised by multiple decay chains can be written: in particular we propose an original method to match final particle spin states among different decay chains able to properly take into account the definition of spin states. For the sake of clarity, we consider a three-body decay $A\to 1,2,3$, but the method presented to write helicity amplitudes is applicable to any decay topology.

Decay amplitudes for multi-body particle decays are obtained in the helicity formalism by breaking the decay chain in sequential two-body decays mediated by intermediate states, for instance a three-body decay is treated by breaking it into two binary decays. Three decay chains, involving three kind of intermediate states, are possible: $A\to R(\to 1,2),3$, $A\to S(\to 1,3),2$ and $A\to U(\to 2,3),1$.

We first consider the $A\to R(\to 1,2),3$ decay chain: the $A\to R,3$ decay can be expressed in the $A$ rest frame by Eq.~\eqref{eq:two_body_amplitude_corrected},
\begin{align}
\mathcal{A}^{A\to R,3}_{m_A,\lambda_R,\bar{\lambda}^R_3}(\theta_R,\phi_R)&= \Braket{p^A_R,\theta_R,\phi_R, \lambda_R,\bar{\lambda}^R_3|\hat{T}|s_A,m_A} \nonumber\\
&= \mathcal{H}^{A\to R,3}_{\lambda_R,\bar{\lambda}^R_3} D^{*s_A}_{m_A,\lambda_R+\bar{\lambda}^R_3}(\phi_R,\theta_R,0),
\end{align}
and the $R\to 1,2$ decay can be written in the same form, in the $R$ rest frame, by applying Eq.~\eqref{eq:two_body_amplitude_corrected} to the $R$ state $\Ket{s_R,\lambda_R}$ as decaying particle,
\begin{align}
\mathcal{A}^{R\to 1,2}_{\lambda_R,\lambda^R_1,\bar{\lambda}^R_2}(\theta^R_1,\phi^R_1)&= \Braket{p^R_1,\theta^R_1,\phi^R_1, \lambda^R_1,\bar{\lambda}^R_2|\hat{T}|\bm{p}^A_R,s_R,\lambda_R} \nonumber\\
&= \mathcal{H}^{R\to 1,2}_{\lambda^R_1,\bar{\lambda}^R_2} D^{*s_R}_{m_R,\lambda^R_1+\bar{\lambda}^R_2}(\phi^R_1,\theta^R_1,0).
\end{align}
The $R$ superscript is put on helicity values and angles of particles 1,2 to stress that their definition is specific to the $A\to R(\to 1,2),3$ decay chain.

The total amplitude of the $A\to R(\to 1,2),3$ decay is written introducing $R$ as intermediate state, and summing the amplitudes over the helicity values $\lambda_R$ satisfying the angular momentum conservation requirements Eq.~\eqref{eq:allowed_helicity_couplings},
\begin{align}
\mathcal{A}&^{A\to R,3 \to 1,2,3}_{m_A,\lambda^R_1,\bar{\lambda}^R_2,\bar{\lambda}^R_3}(\Omega) = \Braket{\lbrace\bm{p}_i\rbrace, \lambda^R_1,\bar{\lambda}^R_2,\bar{\lambda}^R_3|\hat{T}|s_A,m_A}\nonumber\\
&= \sum_{\lambda_R} \Braket{p^R_1,\theta^R_1,\phi^R_1, \lambda^R_1,\bar{\lambda}^R_2|\hat{T}|\bm{p}^A_R,s_R,\lambda_R} \nonumber\\
& \hspace{18pt} \times\Braket{p^A_R,\theta_R,\phi_R, \lambda_R,\bar{\lambda}^R_3|\hat{T}|s_A,m_A}\nonumber\\
&= \sum_{\lambda_R} \mathcal{A}^{A\to R,3}_{m_A,\lambda_R,\bar{\lambda}^R_3}(\theta_R,\phi_R) \mathcal{A}^{R\to 1,2}_{\lambda_R,\lambda^R_1,\bar{\lambda}^R_2}(\theta^R_1,\phi^R_1) \nonumber\\
&= \sum_{\lambda_R}\mathcal{H}^{R\to 1,2}_{\lambda^R_1,\bar{\lambda}^R_2} D^{*s_R}_{\lambda_R,\lambda^R_1+\bar{\lambda}^R_2}(\phi^R_1,\theta^R_1,0)\nonumber\\
& \hspace{18pt} \times \mathcal{H}^{A\to R,3}_{\lambda_R,\bar{\lambda}^R_3} D^{*s_A}_{m_A,\lambda_R+\bar{\lambda}^R_3}(\phi_R,\theta_R,0).
\label{eq:helicity_amplitude_3body_R_state}
\end{align}
Note that the angles entering the decay amplitude depend on the phase space variables describing the decay, denoted collectively as $\Omega$.

Now, let's consider the $A\to S(\to 1,3),2$ decay chain. Its associated amplitude is, following Eq.~\eqref{eq:helicity_amplitude_3body_R_state},
\begin{align}
\mathcal{A}^{A\to S,2 \to 1,2,3}_{m_A,\lambda^S_1,\bar{\lambda}^S_2,\bar{\lambda}^S_3}(\Omega) &= \sum_{\lambda_S}\mathcal{H}^{S\to 1,3}_{\lambda^S_1,\bar{\lambda}^S_3} D^{*s_S}_{\lambda_S,\lambda^S_1+\bar{\lambda}^S_3}(\phi^S_1,\theta^S_1,0)\nonumber\\
& \hspace{14pt} \times\mathcal{H}^{A\to S,2}_{\lambda_S,\bar{\lambda}^S_2} D^{*s_A}_{m_A,\lambda_S+\bar{\lambda}^S_2}(\phi_S,\theta_S,0).
\label{eq:helicity_amplitude_3body_S_state}
\end{align}
Helicity values and angles denoted by the $S$ superscripts are defined specifically for the $A\to S(\to 1,3),2$ decay chain: the definition of final particle spin states for this decay chain is different from that used for the $R$ intermediate state one.

To write the total amplitude of the $A\to 1,2,3$ decay, amplitudes associated to different intermediate states must be summed coherently to properly include interference effects. The sum can be performed only if the definition of final particle spin states is the same across different decay chains. Since helicity systems are specific to each decay chain, they must be rotated to a reference set of spin states, for each final particle. Various solutions to match final particle spin states have been proposed~\cite{Mizuk:2008me,LHCb-PAPER-2015-029,Chen:2017gtx,Mikhasenko:2019rjf}, but none addressed the problem in full generality for generic multi-body decays. In the following we derive the correct matching of final particle spin states requiring that, for any decay chain, final particle states are defined by the same Lorentz transformations relatively to the decaying particle spin states.

The definition of the helicity states used to express the helicity amplitudes Eqs.~\eqref{eq:helicity_amplitude_3body_R_state} and~\eqref{eq:helicity_amplitude_3body_S_state}, is given relatively to the $A$ particle spin states ($S_A$ reference system) by a sequence of Lorentz transformations. Once a conventional definition of the $\ket{s_A,m_A}$ states overall phase is chosen, see Sect.~\ref{sec:phase_definition}, the helicity states are fully specified by the Lorentz transformation sequence, overall phase included. Therefore, to relate different helicity state definitions it is mandatory to refer back to the initial $S_A$ reference system, \textit{i.e.} it is not possible to relate the two systems via a direct transformation. To stress this essential point, let's consider the two helicity systems for particle 1 defined by the $R$ and $S$ decay chains, $S_1^{HH,R}$ and $S_1^{HH,S}$, respectively. Suppose we find a rotation $\hat{R}$ such that $\hat{R}S_1^{HH,S} = S_1^{HH,R}$ and we rotate the $\ket{\bm{p}^S_1,s_1,\lambda^S_1}$ states applying the Wigner $D$-matrix associated to that rotation. However, this does not guarantee that the spin state phase definition is the same between $S_1^{HH,R}$ and $\hat{R}S_1^{HH,S}$ systems, since they are defined with respect to $S_A$ by different Lorentz transformation sequences: for a fermion, the two may differ by an overall $2\pi$ rotation changing the relative sign among spin states, by Eq.~\ref{eq:2pi_rotation_fermion}.

The correct way to proceed is to define a reference spin system for the final particle from the initial system $S_A$ and relate each helicity system to this one by applying a sequence of rotations\footnote{The reference and the helicity systems are different rest frames of the final particle, so no boosts need to be applied.} which turn the Lorentz transformation sequence defining the helicity state into that specifying the reference one. Any spin system can be chosen as reference one, we will choose canonical states reached from $S_A$ being the simplest possibility.

We illustrate the method explicitly deriving the rotation sequences transforming the helicity systems of the $A\to 1,2,3$ decay final particles, for $R$ and $S$ intermediate state decay chains, to their canonical systems reached from $S_A$. Starting from particle 1, the helicity system $S_1^{HH,R}$ is defined from $S_A$ by the sequence, see Eq.~\eqref{eq:helicity_state_definition},
\begin{align}
S_R^H &= L(-p^A_R \bm{z})R(0,-\theta_R,-\phi_R) S_A \nonumber\\
S_1^{HH,R} &= L(-p^R_1 \bm{z})R(0,-\theta^R_1,-\phi^R_1) S_R^H.
\label{eq:three_body_helicity_states}
\end{align}
in which $S_R^H$ is the $R$ helicity system reached from $S_A$.
The canonical system of particle 1 derived from $S_R^{H}$, following Eq.~\eqref{eq:canoncial_state_definition} by decomposing the Lorentz boost, is
\begin{align}
S_1^{HC,R} &= R(\phi^R_1,\theta^R_1,0) L(-p^R_1 \bm{z})R(0,-\theta^R_1,-\phi^R_1) S_R^H\nonumber\\
&= R(\phi^R_1,\theta^R_1,0) S_1^{HH,R},
\label{eq:three_body_canhel_states}
\end{align}
so that, the rotation $R(\phi^R_1,\theta^R_1,0)$ from $S_1^{HH,R}$ to $S_1^{HC,R}$ systems ``undoes'' the helicity rotation $R(0,-\theta^R_1,-\phi^R_1)$, acting in the particle 1 rest frame. The spin states $\Ket{\bm{p}^R_1,s_1,\mu^R_1}$ defined by the $S_1^{HC,R}$ system Lorentz transformation are expressed in terms of the helicity states $\Ket{\bm{p}^R_1,s_1,\lambda^R_1}$ as in Eq.~\eqref{eq:canonical_helicity_relation_states},
\begin{equation}
\hat{R}(\phi^R_1,\theta^R_1,0)\Ket{\bm{p}^R_1,s_1,\lambda^R_1} = \sum_{\mu^R_1} D^{s_1}_{\mu^R_1,\lambda^R_1}(\phi^R_1,\theta^R_1,0) \Ket{\bm{p}^R_1,s_1,\mu^R_1}.
\label{eq:particle_1_helicity1_rotation}
\end{equation}
Now, let's consider the canonical system of particle 1 $S_1^{CC,R}$ derived from the $R$ canonical system $S_R^C$, defined following Eq.~\eqref{eq:canoncial_state_definition},
\begin{align}
S_R^C &= L(-\bm{p}^A_R) S_A \nonumber\\
&= R(\phi_R,\theta_R,0)L(-p^A_R \bm{z})R(0,-\theta_R,-\phi_R) S_A \nonumber\\
S_1^{CC,R} &= L(-\bm{p}'^R_1) S_R^C \nonumber\\
&= R(\phi'^R_1,\theta'^R_1,0)L(-p^R_1 \bm{z})R(0,-\theta'^R_1,-\phi'^R_1) S_R^C,
\label{eq:three_body_canonical_states}
\end{align}
in which primed quantities indicate that the particle 1 momentum used to build the $S_1^{CC,R}$ system is different from the one defining the $S_1^{HC,R}$ system, being determined from $S_R^C$ instead of $S_R^H$. The momenta $\bm{p}^R_1$ and $\bm{p}'^R_1$ differ by the additional rotation $R(\phi_R,\theta_R,0)$ used in the definition of $S_R^C$,
\begin{equation}
\bm{p}'^R_1 = R(\phi_R,\theta_R,0)\bm{p}^R_1,
\end{equation}
and the rotation $R(0,-\theta'^R_1,-\phi'^R_1)$ used to align $\bm{p}'^R_1$ to the $z$ axis can be decomposed into $R(0,-\theta_R,-\phi_R)$, rotating $\bm{p}'^R_1$ into $\bm{p}^R_1$, times $R(0,-\theta^R_1,-\phi^R_1)$, the helicity rotation aligning $\bm{p}^R_1$ with the $z$ axis,
\begin{equation}
R(0,-\theta'^R_1,-\phi'^R_1) = R(0,-\theta^R_1,-\phi^R_1) R(0,-\theta_R,-\phi_R).
\end{equation}
Applying the above decomposition, along with its inverse,
\begin{equation}
R(\phi'^R_1,\theta'^R_1,0) = R(\phi_R,\theta_R,0) R(\phi^R_1,\theta^R_1,0),
\end{equation}
to Eq.~\eqref{eq:three_body_canonical_states} and using the systems defined in Eqs.~\eqref{eq:three_body_helicity_states} and \eqref{eq:three_body_canhel_states}, we obtain
\begin{align}
S_1^{CC,R} &= L(-\bm{p}'^R_1) L(-\bm{p}^A_R) S_A \nonumber\\
&= R(\phi'^R_1,\theta'^R_1,0)L(-p^R_1 \bm{z})R(0,-\theta'^R_1,-\phi'^R_1) \nonumber\\
&\times R(\phi_R,\theta_R,0)L(-p^A_R \bm{z})R(0,-\theta_R,-\phi_R) S_A \nonumber\\
&= R(\phi'^R_1,\theta'^R_1,0)L(-p^R_1 \bm{z})R(0,-\theta^R_1,-\phi^R_1) \nonumber\\
&\times L(-p^A_R \bm{z})R(0,-\theta_R,-\phi_R) S_A \nonumber\\
&= R(\phi_R,\theta_R,0) R(\phi^R_1,\theta^R_1,0) S_1^{HH,R} \nonumber\\
&= R(\phi_R,\theta_R,0) S_1^{HC,R}.
\end{align}
The rotation $R(\phi_R,\theta_R,0)$ from $S_1^{HC,R}$ to $S_1^{CC,R}$ systems ``undoes'' the helicity rotation $R(0,-\theta_R,-\phi_R)$, still acting in the particle 1 rest frame. The spin states $\Ket{\bm{p}'^R_1,s_1,\nu^R_1}$ defined by the $S_1^{CC,R}$ system are expressed in terms of the states $\Ket{\bm{p}^R_1,s_1,\mu^R_1}$ as
\begin{equation}
\hat{R}(\phi_R,\theta_R,0)\Ket{\bm{p}^R_1,s_1,\mu^R_1} = \sum_{\nu^R_1} D^{s_1}_{\nu^R_1,\mu^R_1}(\phi_R,\theta_R,0) \Ket{\bm{p}'^R_1,s_1,\nu^R_1}.
\label{eq:particle_1_helicityR_rotation}
\end{equation}
Finally, the system $S_1^{CC,R}$ has to be related to our reference system, the canonical system of particle 1 reached directly from $S_A$,
\begin{equation}
S_1^{C,A} = L(-\bm{p}^A_1) S_A,
\end{equation}
which differs from $S_1^{CC,R}$ by
\begin{equation}
S_1^{C,A} = L(-\bm{p}^A_1) L(\bm{p}^A_R) L(\bm{p}'^R_1) S_1^{CC,R}.
\end{equation}
The transformation $L(-\bm{p}^A_1) L(\bm{p}^A_R) L(\bm{p}'^R_1)$ is equivalent to a rotation, called Wigner rotation, indicated generically as an Euler rotation $R(\alpha^{W,R}_1,\beta^{W,R}_1,\gamma^{W,R}_1)$, in which the associated Euler angles do not have a simple algebraic expression, but can be easily computed\footnote{The value of the Euler angles can be calculated as follows: the Wigner rotation matrix is found from multiplying the three matrices associated to each Lorentz boost; the unit vectors describing the rotated system are simply the columns of the Wigner rotation matrix; finally Euler angles are computed using Eq.~\eqref{eq:euler_angles}. Explicit axis-angle expressions for the composition of two Lorentz boosts are reported in Ref.~\cite{Gourgoulhon}.}.
The canonical spin states $\Ket{\bm{p}^A_1,s_1,m_1}$ defined by the $S_1^{C,A}$ coordinate systems are expressed in terms of the states $\Ket{\bm{p}'^R_1,s_1,\nu^R_1}$ as
\begin{align}
\hat{R}(\alpha^{W,R}_1,&\beta^{W,R}_1,\gamma^{W,R}_1)\Ket{\bm{p}'^R_1,s_1,\nu^R_1}= \sum_{m_1} D^{s_1}_{m_1,\nu^R_1}(\alpha^{W,R}_1,\beta^{W,R}_1,\gamma^{W,R}_1) \Ket{\bm{p}^A_1,s_1,m_1}.
\label{eq:particle_1_wigner_rotation}
\end{align}

\tikzstyle{int}=[draw, minimum size=2em]
\tikzstyle{init} = [pin edge={to-,thin,black}]

\newcommand\scale{1.8}

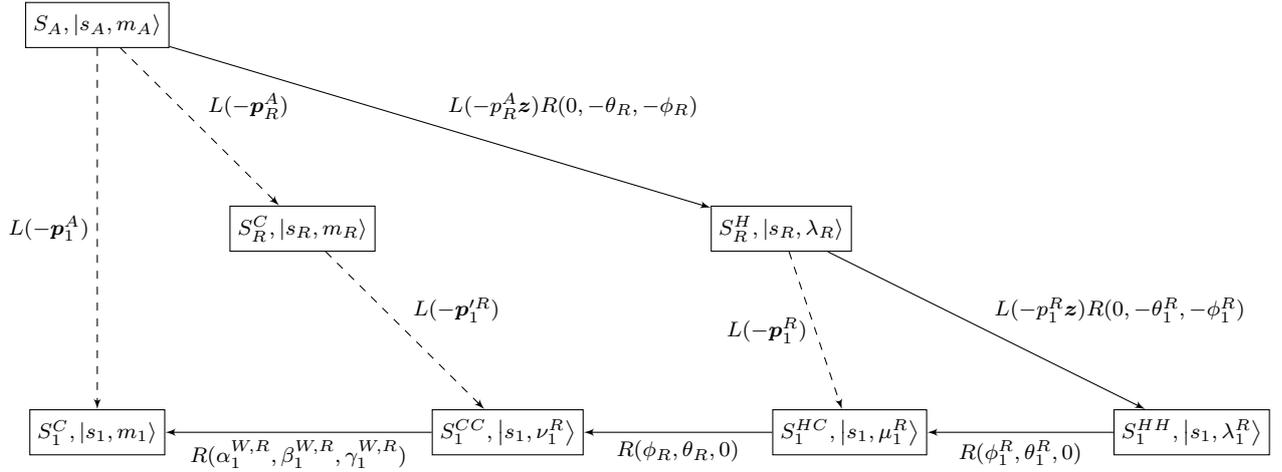
\begin{figure*}
\centering
\begin{tikzpicture}[node distance=2*\scale cm,auto,>=latex']
\begin{scriptsize}
    \node [int] (a1) {$S_A,\Ket{s_A,m_A}$};
    \node (a2) [right of=a1,node distance=1.5*\scale cm] {};
    \node [int] (b1) [below of=a2,node distance=1.5*\scale cm] {$S^C_R,\Ket{s_R,m_R}$};
    \node [int] (b2) [right of=b1, node distance=3.5*\scale cm] {$S^H_R,\Ket{s_R,\lambda_R}$};
    \node [int] (c1) [below of=a1, node distance=3*\scale cm]{$S^C_1,\Ket{s_1,m_1}$};
    \node [int] (c2) [right of=c1, node distance=3*\scale cm]{$S^{CC}_1,\Ket{s_1,\nu^R_1}$};
    \node [int] (c3) [right of=c2, node distance=2.5*\scale cm]{$S^{HC}_1,\Ket{s_1,\mu^R_1}$};
    \node [int] (c4) [right of=c3, node distance=2.5*\scale cm]{$S^{HH}_1,\Ket{s_1,\lambda^R_1}$};
    \path[->,left,dashed] (a1) edge node {$L(-\bm{p}^A_1)$} (c1);
    \path[->,dashed] (a1) edge node {$L(-\bm{p}^A_R)$} (b1);
    \path[->] (a1) edge node {$L(-p^A_R \bm{z})R(0,-\theta_R,-\phi_R)$} (b2);
    \path[->,dashed] (b1) edge node {$L(-\bm{p}'^R_1)$} (c2);
    \path[->,left,dashed] (b2) edge node {$L(-\bm{p}^R_1)$} (c3);
    \path[->] (b2) edge node {$L(-p^R_1 \bm{z})R(0,-\theta^R_1,-\phi^R_1)$} (c4);
    \path[->] (c4) edge node {$R(\phi^R_1,\theta^R_1,0)$} (c3);
    \path[->] (c3) edge node {$R(\phi_R,\theta_R,0)$} (c2);
    \path[->] (c2) edge node {$R(\alpha^{W,R}_1,\beta^{W,R}_1,\gamma^{W,R}_1)$} (c1);        
\end{scriptsize}
\end{tikzpicture}
\caption{Graphical summary of the spin state definitions associated to the decay amplitude for final particle 1 and intermediate state $R$. The solid lines indicate the Lorentz transformation sequence involved in the writing of the helicity amplitude. Once a conventional choice for $\Ket{s_A,m_A}$ initial $A$ particle spin states is set, all the other spin states are completely specified by their Lorentz transformation sequence. Momentum labels of spin states have been suppressed to avoid cluttering the notation.\label{fig:diagram}}
\end{figure*}

The different Lorentz transformations applied starting from the initial $S_A$ system down to the final $S_1^{C,A}$ one, for $R$ intermediate states, are graphically summarised in Fig.~\ref{fig:diagram}, along with the definition of the many spin states involved in the writing of the helicity amplitude.

Combining Eqs.~\eqref{eq:particle_1_helicity1_rotation},~\eqref{eq:particle_1_helicityR_rotation},~\eqref{eq:particle_1_wigner_rotation}, the relation between particle 1 canonical spin states defined by the $S_1^{C,A}$ system and particle 1 helicity states defined by the $S_1^{HH,R}$ system, those expressing the decay amplitude Eq.~\eqref{eq:helicity_amplitude_3body_R_state}, is
\begin{align}
&\hat{R}(\alpha^{W,R}_1,\beta^{W,R}_1,\gamma^{W,R}_1) \hat{R}(\phi_R,\theta_R,0) \hat{R}(\phi^R_1,\theta^R_1,0)\Ket{\bm{p}^R_1,s_1,\lambda^R_1}\nonumber\\
&= \sum_{\mu^R_1} D^{s_1}_{\mu^R_1,\lambda^R_1}(\phi^R_1,\theta^R_1,0)\nonumber\\
&\times \sum_{\nu^R_1} D^{s_1}_{\nu^R_1,\mu^R_1}(\phi_R,\theta_R,0)\nonumber\\
&\times \sum_{m_1} D^{s_1}_{m_1,\nu^R_1}(\alpha^{W,R}_1,\beta^{W,R}_1,\gamma^{W,R}_1) \Ket{\bm{p}^A_1,s_1,m_1}
\label{eq:final_particle_spin_rotation}
\end{align}

The transformation sequence for particle 2 is almost identical to particle 1, but for the Lorentz boost involving its momentum, thanks to the use of the opposite-helicity states we introduced in Sect.~\ref{sec:helicity_formalism_rev}. The sequence of helicity rotations is indeed the same as for particle 1, since particle 2 opposite-helicity states are defined by the same helicity rotation specifying particle 1 helicity states, see Eq.~\eqref{eq:canonical_opposite_helicity_relation}. The relation between particle 2 canonical spin states ($S_2^{C,A}$ system) and particle 2 opposite-helicity states ($S_2^{OHH,S}$ system) is thus the same as Eq.~\eqref{eq:final_particle_spin_rotation} apart from the Wigner rotation
\begin{equation}
R(\alpha^{W,R}_2,\beta^{W,R}_2,\gamma^{W,R}_2) = L(-\bm{p}^A_2) L(\bm{p}^A_R) L(\bm{p}'^R_2).
\end{equation}
For particle 2 the advantage of using our proposed definition for two-particle product helicity states Eq.~\eqref{eq:plane_wave_helicity_corrected} instead of the usual one Eq.~\eqref{eq:plane_wave_helicity} is evident: in the latter case two rotations must be performed, one corresponding to the helicity rotation and one for the inversion.

For particle 3, the spin rotation to its canonical system is much simpler since only one helicity rotation has to be ``undone'', with no Wigner rotation needed since a single direct boost from particle $A$ to particle 3 rest frames is involved, leading to
\begin{equation}
\Ket{\bm{p}^A_3,s_3,\lambda^R_3} = \sum_{m_3} D^{s_3}_{m_3,\lambda^R_3}(\phi_R,\theta_R,0) \Ket{\bm{p}^A_3,s_3,m_3}.
\end{equation}

The final particle spin rotations are introduced in the decay amplitudes as follows. The amplitude for $R$ intermediate states we are interested in is
\begin{equation}
\mathcal{A}^{A\to R,3 \to 1,2,3}_{m_A,m_1,m_2,m_3}(\Omega) = \Braket{\lbrace\bm{p}_i\rbrace, m_1,m_2,m_3|\hat{T}|s_A,m_A},
\label{eq:helicity_amplitude_3body_R_state_definition}
\end{equation}
in which the final particles state
\begin{align}
\Ket{\lbrace\bm{p}_i\rbrace, m_1,m_2,m_3} = \Ket{s_1,m_1} \otimes \Ket{s_2,m_2} \otimes \Ket{s_3,m_3}
\label{eq:final_particle_rotated_state}
\end{align}
is the product of the canonical spin states reached from the $S_A$ system for each final particle.

For the sake of clarity, we consider a much simpler example: the case of a single rotation on a single particle state. Suppose we know the transition amplitude $\mathcal{A}_{\lambda} = \Braket{s,\lambda|\hat{T}| i}$ between a given initial state $\Ket{i}$ and a final state $\Ket{s,\lambda}$, but we would like to express the amplitude with respect to some rotated spin states
\begin{equation}
\hat{R}(\alpha,\beta,\gamma)\Ket{s,\lambda} = \sum_m D^s_{m,\lambda}(\alpha,\beta,\gamma)\Ket{s,m}.
\end{equation}
To this end, we write the transition amplitude for the rotated states introducing the $\Ket{s,\lambda}$ states as a set of intermediate states,
\begin{align}
\mathcal{A}_{m} &= \Braket{s,m|\hat{T}| i}\nonumber\\
&= \sum_{\lambda} \Braket{s,m|\hat{R}(\alpha,\beta,\gamma)|s,\lambda} \Braket{s,\lambda|\hat{T}| i}\nonumber\\
&= \sum_{\lambda} D^s_{m,\lambda}(\alpha,\beta,\gamma) \mathcal{A}_{\lambda},
\end{align}
in which the rotation operator $\hat{R}(\alpha,\beta,\gamma)$ acts as transition operator between final particle spin states. Note the parallel with Eq.~\eqref{eq:two_body_amplitude_rewritten}: the Wigner $D$-matrix is indeed the representation of the rotation applied.

Generalising the case of a single rotation for a single particle, the amplitude for $R$ intermediate states Eq.~\eqref{eq:helicity_amplitude_3body_R_state_definition}, expressed for the final particles canonical state Eq.~\eqref{eq:final_particle_rotated_state} in terms of the amplitude for helicity states Eq.~\eqref{eq:helicity_amplitude_3body_R_state},
becomes
\begin{align}
\mathcal{A}^{A\to R,3 \to 1,2,3}_{m_A,m_1,m_2,m_3}(\Omega) &= \hspace{-6pt}\sum_{\lambda^R_1,\mu^R_1,\nu^R_1} 
D^{s_1}_{m_1,\nu^R_1}(\alpha^{W,R}_1,\beta^{W,R}_1,\gamma^{W,R}_1) \, D^{s_1}_{\nu^R_1,\mu^R_1}(\phi_R,\theta_R,0) \, D^{s_1}_{\mu^R_1,\lambda^R_1}(\phi^R_1,\theta^R_1,0) \nonumber\\
&\times \hspace{-6pt}\sum_{\lambda^R_2,\mu^R_2,\nu^R_2}
D^{s_2}_{m_2,\nu^R_2}(\alpha^{W,R}_2,\beta^{W,R}_2,\gamma^{W,R}_2) \, D^{s_2}_{\nu^R_2,\mu^R_2}(\phi_R,\theta_R,0) \, D^{s_2}_{\mu^R_2,\lambda^R_2}(\phi^R_1,\theta^R_1,0) \nonumber\\
&\times \sum_{\lambda^R_3} 
D^{s_3}_{m_3,\lambda^R_3}(\phi_R,\theta_R,0) \nonumber\\
&\times \mathcal{A}^{A\to R,3 \to 1,2,3}_{m_A,\lambda^R_1,\bar{\lambda}^R_2,\bar{\lambda}^R_3}(\Omega).
\label{eq:helicity_amplitude_3body_R_state_rotated}
\end{align}

The amplitude for $S$ intermediate states can be written analogously, taking into account the different decay topology (interchanging particle 2 and 3 role),
\begin{align}
\mathcal{A}^{A\to S,2 \to 1,2,3}_{m_A,m_1,m_2,m_3}(\Omega) &= \hspace{-6pt}\sum_{\lambda^S_1,\mu^S_1,\nu^S_1}
D^{s_1}_{m_1,\nu^S_1}(\alpha^{W,S}_1,\beta^{W,S}_1,\gamma^{W,S}_1) \, D^{s_1}_{\nu^S_1,\mu^S_1}(\phi_S,\theta_S,0) \, D^{s_1}_{\mu^S_1,\lambda^S_1}(\phi^S_1,\theta^S_1,0) \nonumber\\
&\times \sum_{\lambda^S_2}
D^{s_2}_{m_2,\lambda^S_2}(\phi_S,\theta_S,0) \nonumber\\
&\times \hspace{-6pt}\sum_{\lambda^S_3,\mu^S_3,\nu^S_3}
D^{s_3}_{m_3,\nu^S_3}(\alpha^{W,S}_3,\beta^{W,S}_3,\gamma^{W,S}_3) \, D^{s_3}_{\nu^S_3,\mu^S_3}(\phi_S,\theta_S,0) \, D^{s_3}_{\mu^S_3,\lambda^S_3}(\phi^S_1,\theta^S_1,0) \nonumber\\
&\times \mathcal{A}^{A\to S,2 \to 1,2,3}_{m_A,\lambda^S_1,\bar{\lambda}^S_2,\bar{\lambda}^S_3}(\Omega).
\label{eq:helicity_amplitude_3body_S_state_rotated}
\end{align}

Finally, the total decay amplitude is obtained summing the amplitudes associated to each intermediate state,
\begin{align}
\mathcal{A}^{A\to 1,2,3}_{m_A,m_1,m_2,m_3}(\Omega) &=
\sum_i \mathcal{A}^{A\to R_i,3 \to 1,2,3}_{m_A,m_1,m_2,m_3}(\Omega) \nonumber\\
&+ \sum_j \mathcal{A}^{A\to S_j,2 \to 1,2,3}_{m_A,m_1,m_2,m_3}(\Omega) \nonumber\\
&+ \sum_k \mathcal{A}^{A\to U_k,1 \to 1,2,3}_{m_A,m_1,m_2,m_3}(\Omega),
\label{eq:helicity_amplitude_3body_summed}
\end{align}
in which the amplitude for $U$ intermediate states can be written similarly as for $R$ and $S$ states.

For a generic multi-body decay the generalisation of the method for the matching of the final particle spin states is straightforward. Each final particle helicity state must be rotated ``undoing'' all the helicity rotations applied along its decay chain, in reversed order. One additional Wigner rotation is needed, describing the difference between the direct boost defining the particle canonical state and the boost sequence applied along its decay chain.


\section{Effects of an incorrect matching of final particles spin states on decay distributions}
\label{sec:effects_phase}
We stress the need for a correct matching of final particle spin states by considering the consequences an incorrect matching can have. We first present a general discussion of the observable effects one can introduce in the decay distributions even in the simple case the incorrect matching just introduces a relative phase among spin states belonging to different decay chains. Then, we present a numerical study in which we compare the method for the matching of final particle spin states presented in this article to the one employed in Refs.~\cite{Mizuk:2008me,LHCb-PAPER-2015-029}: we show the latter is not correct since breaks rotational invariance.

We first discuss the general case of a particle $A$ decay to $\lbrace i=1,..,n \rbrace$ final state particles, passing through two intermediate states $R$ and $S$. The associated amplitudes are denoted as $\mathcal{A}^R_{m_A,\lbrace m_i\rbrace}$ and $\mathcal{A}^S_{m_A,\lbrace m_i\rbrace}$. Let's suppose that, due to an incorrect matching of final particle spin state definition, a phase difference $\exp\left[i\psi(\Omega)\right]$ is introduced between the two amplitudes. Since the definition of the spin systems depends on the phase space variables, the incorrect phase will be in general function of the decay phase space. The effects of such a phase are in a certain sense analogous to the phase effects seen in interferometry experiments: in that case, the phase difference between particles following different paths is a physical effect produced by a difference in energy potential felt by the particle; in our case, it is an unnatural effect caused by an incorrect definition of spin states among different decay chains (the analogue of paths). Also the mere change of sign of fermion states under $2\pi$ angle rotations is observable, being measured in neutron interferometry~\cite{RAUCH1975425,Werner:1975wf}.

The polarised decay rate for definite initial $m_A$ and final spin projections $\lbrace m_i\rbrace$, see Eq.~\eqref{eq:decay_rate_pure_states} of Appendix~\ref{sec:polarised_rate}, is
\begin{align}
p_{m_A,\lbrace m_i\rbrace}(\Omega) &= \left| \mathcal{A}^R_{m_A,\lbrace m_i\rbrace} + \exp\left[i\psi(\Omega)\right]\mathcal{A}^S_{m_A,\lbrace m_i\rbrace} \right|^2 \nonumber\\
&= \left| \mathcal{A}^R_{m_A,\lbrace m_i\rbrace} \right|^2 + \left| \mathcal{A}^S_{m_A,\lbrace m_i\rbrace} \right|^2 + 2 Re\left[\exp\left[-i\psi(\Omega)\right] \mathcal{A}^R_{m_A,\lbrace m_i\rbrace} \mathcal{A}^{S*}_{m_A,\lbrace m_i\rbrace} \right].
\label{eq:decay_rate_phase}
\end{align}
The incorrect phase affects the decay rate modifying the interference terms between different decay chains, changing their functional form in terms of phase space variables. In an amplitude fit to experimental data, it means that fit parameters depending on interference effects between different decay chains can be biased due to the incorrect phase space dependence introduced. Note that the incorrect phase also affects the unpolarised decay rate: there is no guarantee that the sum over the spin states mitigates its effect.

A numerical study illustrating these kind of unnatural effects is performed on the decay distributions of the three-body \Lcpkpi decay obtained from its helicity amplitudes. The three-body decay phase space is described by 5 degrees-of-freedom: two ``Dalitz'' two-body invariant masses and three angles describing the orientation of the decay with respect to the $A$ reference system $S_A$, called orientation angles in the following. For the \Lcpkpi decay they can be chosen to be $\mqpk$, $\mqkpi$, the cosine of the proton polar angle in the $S_\Lc$ system, $\cos\theta_p$, the proton azimuthal angle, $\phi_p$, and the signed angle between the plane formed by the proton and the \Lc quantisation axis and the plane formed by the kaon and the pion, named $\chi$.

We test a property of the decay distributions following from rotational invariance, which must be satisfied irrespective of the amplitude model considered: for zero \Lc polarisation the orientation angles distributions $\cos\theta_p,\phi_p,\chi$ must be uniform. Indeed, in absence of a polarisation vector, nothing specifies a direction in the $S_{\Lc}$ system. This property provides a necessary test for the correctness of an amplitude model: if the model produces anisotropic orientation angle distributions for zero polarisation it is wrong, violating rotational invariance.

We consider \Lcpkpi helicity amplitudes written closely following the three-body ones described in Sect.~\ref{sec:helicity_amplitudes} (details specific to the \Lcpkpi decay are reported in Appendix~\ref{sec:Lcpkpi_model}) applying two different methods for matching proton spin states among different decay chains: the one presented in this article and the one employed for the amplitude analyses Refs.~\cite{Mizuk:2008me,LHCb-PAPER-2015-029}. For the \Lcpkpi case, the latter method consists in applying a single rotation to the proton states aligning their quantisation axes to a reference one, which are defined as the direction opposite to the momentum of the particle recoiling against the proton in the proton rest frame. Taking the $K^*(\to K^-\pi^+)$ decay chain as reference, the rotation applied to the proton helicity states is
\begin{align}
\hat{R}(0,\beta^R_p,0) \Ket{\bm{p}^{R}_p,1/2, \lambda_p} &= d^{1/2}_{\lambda'_p,\lambda_p}(\beta^R_p) \Ket{\bm{p}^{R}_p, 1/2, \lambda'_p},
\label{eq:proton_spin_rotation_PQ}
\end{align}
with angles
\begin{align}
\cos\beta^{\Lz^*}_p = \hat{\bm{p}}^p_{K^*} \cdot \hat{\bm{p}}^p_{K^-}, \hspace{24pt}
\cos\beta^{\Deltares^*}_p = \hat{\bm{p}}^p_{K^*} \cdot \hat{\bm{p}}^p_{\pi^+},
\end{align}
for the $\Lz^*(\to pK^-)$ and $\Deltares^{*++}(\to p\pi^+)$ decay chains, respectively.

For the numerical study we consider a \Lcpkpi amplitude model consisting of three resonance states, one per decay channel. The detailed specification of the helicity couplings and resonance descriptions employed are reported in Appendix~\ref{sec:Lcpkpi_model}. The code reproducing the amplitude model is based on a version of the TensorFlowAnalysis package~\cite{TFA} adapted to five-dimensional phase space three-body amplitude fits~\cite{Marangotto:2713231}; this package depends on the machine-learning framework TensorFlow~\cite{tensorflow2015-whitepaper} and the ROOT package~\cite{Brun:1997pa}.

The phase space distributions are described by a set of five millions Monte Carlo pseudo-data generated according to the two amplitude models for zero \Lc polarisation. We stress that the two amplitude models only differ by the proton spin matching method.

\begin{figure}
\centering
\includegraphics[width=\textwidth]{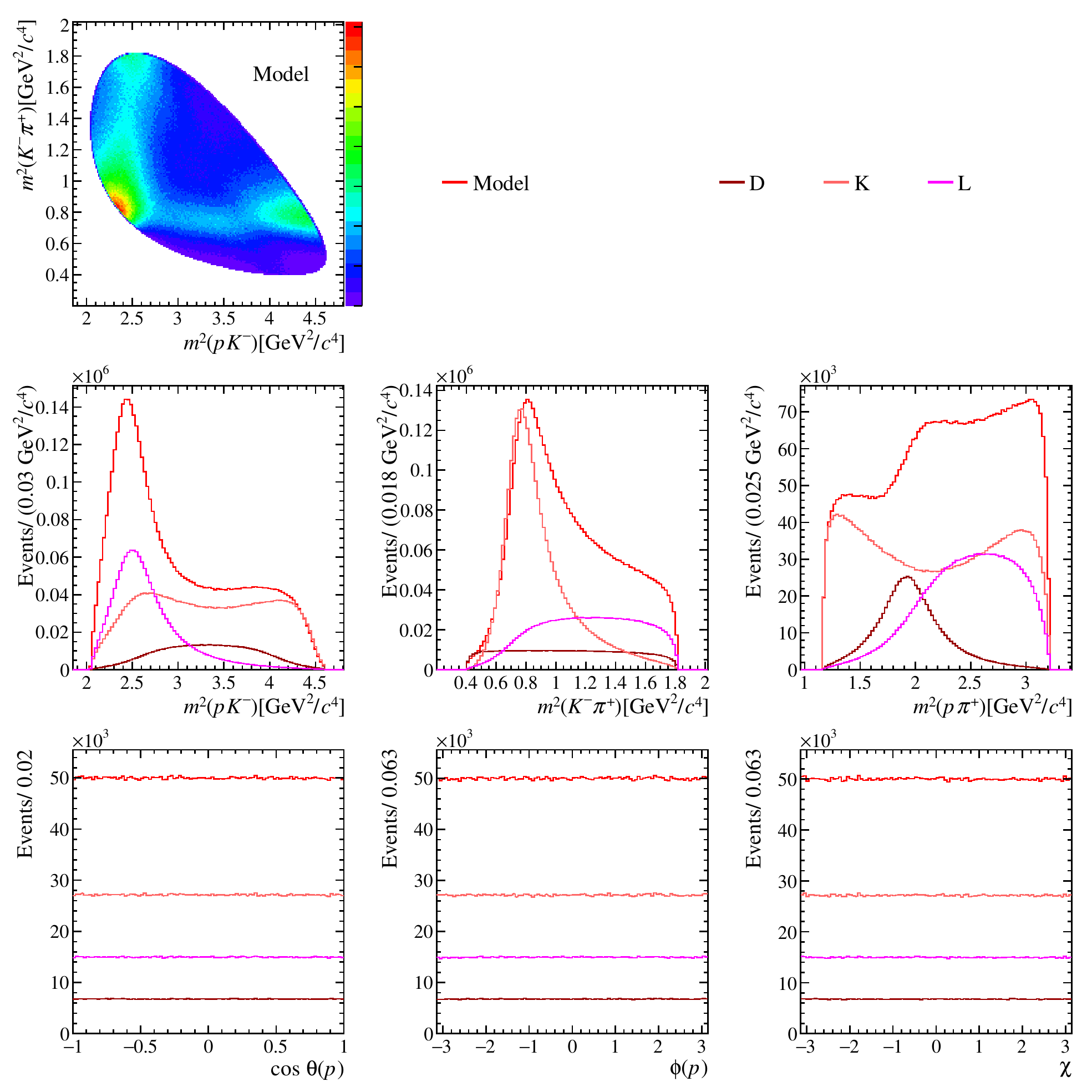}
\caption{Phase space decay distributions for the \Lcpkpi amplitude models written using the proton spin matching method presented in this article. ``Model'' refers to the full \Lcpkpi amplitude model distributions, while ``D'', ``L'', ``K'' labels indicate the distributions associate to single $\Deltares^{*++}$, $\Lz^*$, $K^*$ resonance contributions.\label{fig:Test_iso_pol_frame}}
\end{figure}

The distributions associated to the method presented in this article are shown in Fig.~\ref{fig:Test_iso_pol_frame}: the orientation angle distributions are precisely isotropic, given the accuracy allowed by the large pseudo-data sample. 

\begin{figure}
\centering
\includegraphics[width=\textwidth]{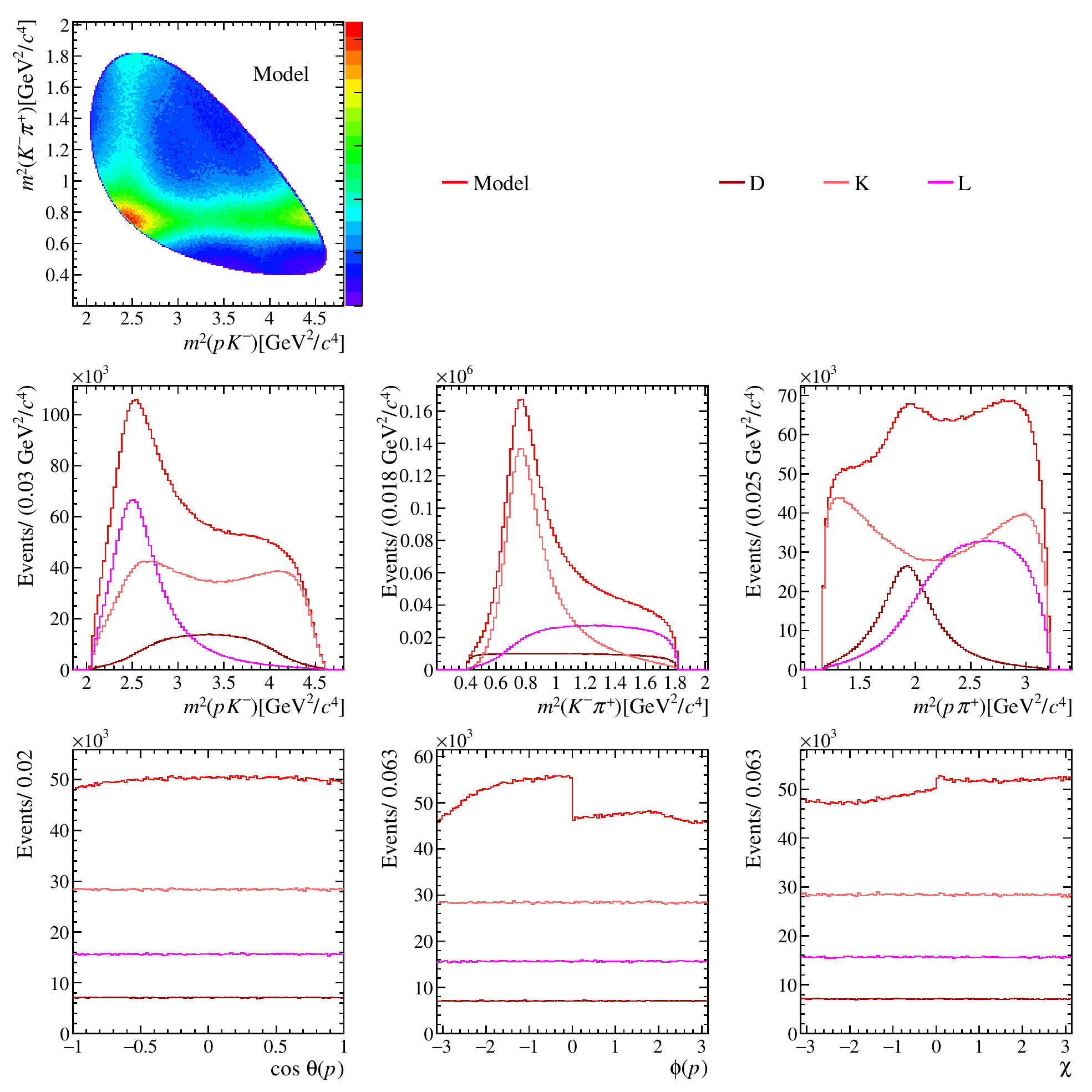}
\caption{Phase space decay distributions for the \Lcpkpi amplitude models written using the proton spin matching method used in Refs.~\cite{Mizuk:2008me,LHCb-PAPER-2015-029}. ``Model'' refers to the full \Lcpkpi amplitude model distributions, while ``D'', ``L'', ``K'' labels indicate the distributions associate to single $\Deltares^{*++}$, $\Lz^*$, $K^*$ resonance contributions.\label{fig:Test_iso_PQ}}
\end{figure}

The distributions associated to the spin matching method used in Refs.~\cite{Mizuk:2008me,LHCb-PAPER-2015-029} are shown in Fig.~\ref{fig:Test_iso_PQ}: the orientation angle distributions are evidently anisotropic, also featuring a step in the $\phi_p$ distribution. This method is thus incorrect, leading to an amplitude model violating rotational invariance. Moreover, it is important to note that the invariant mass distributions are different for the two methods: therefore an incorrect spin matching can also affect Dalitz plot analyses, in which orientation angles are integrated over the decay rate, as in spin-zero meson decays or when the polarisation of the decaying particle is not considered.

To conclude, we discuss why the matching method of Refs.~\cite{Mizuk:2008me,LHCb-PAPER-2015-029} is not correct. The proton spin rotations Eq.~\eqref{eq:proton_spin_rotation_PQ} ensure that the definition of the proton quantisation axis is the same among different decay chains. Therefore, the proton states can differ among different decay chains by a rotation of the orthogonal axes, Eq.~\ref{eq:state_transformation}, or by the sign of the proton state, Eq.~\eqref{eq:2pi_rotation_fermion}, which introduce an incorrect phase difference among decay chains. This is exactly the case considered at the beginning of this section. Indeed, the decay distributions associated to single resonances are the same for the two matching methods (up to an irrelevant normalisation factor), Fig.~\ref{fig:Test_iso_PQ}, since the phase difference $\exp\left[i\psi(\Omega)\right]$ cancels for single resonance decay rates. However, the incorrect phase affects the decay distributions of the full amplitude model, which includes interference effects among different decay chains, as in Eq.~\eqref{eq:decay_rate_phase}.

\section{Conclusions}
\label{sec:conclusions}
We presented the general expression of helicity amplitudes for generic multi-body particle decays characterised by multiple decay chains. We demonstrated the importance of a precise specification of spin states for a correct writing of helicity amplitudes, proposing an original method to match final particle spin states among different decay chains, applicable to any multi-body decay topology.

To this end, we reviewed the definition of spin states in quantum mechanics, with a particular attention to their phase specifications, and the helicity formalism, considering a simpler definition of two-particle helicity states than the standard one. The proposed method to match final particle spin states was obtained requiring that, for any decay chain, final particle states are defined by the same Lorentz transformations relatively to the decaying particle spin states, by applying a sequence of rotations. Helicity amplitudes were explicitly written for three-body decays.

We discussed the consequences of an incorrect phase introduced between amplitudes describing different intermediate states on the decay distributions, showing they produce observable effects on the decay distributions via interference terms. We also tested numerically our spin matching method against the one used for the amplitude analyses Refs.~\cite{Mizuk:2008me,LHCb-PAPER-2015-029}, in the case of \Lcpkpi decay amplitudes. We showed how the latter leads to amplitude models violating rotational invariance while the first does not. This incorrect behaviour follows from a wrong phase introduced between spin states among different decay chains.

\section*{Data Availability}
The articles used to support the findings of this study are
included within the article and are cited at relevant places
within the text as references.

\section*{Conflicts of Interest}
The author declares that there are no conflicts of interest.

\section*{Funding Statement}
This work was supported by the ERC Consolidator Grant SELDOM G.A. 771642.

\section*{Acknowledgements}
I thank Mikhail Mikhasenko and Alessandro Pilloni for the critical review of the first version of the article, in particular for pointing out the need of a Wigner rotation in the matching of final particle spin states. I thank my colleagues Louis Henry, Fernando Mart\'{i}nez Vidal, Andrea Merli, Nicola Neri and Elisabetta Spadaro Norella for interesting discussions about the helicity formalism and its application to baryon decays.

\appendix

\section{Spin in relativistic processes}
\label{sec:spin_relativistic}
Here, we review the description of relativistic processes involving particles with spin, which is complicated by the impossibility of a covariant definition of the spin operators~\cite{Leader2011}. Indeed, a set of spin operators $\bm{\hat{S}}$ is well-defined only when acting on states describing particles at rest, so that a different set of spin operators has to be defined for each particle to express their spin states.

To describe processes involving spins of particles in relative motion, the definition of spin states must be linked to their relative kinematics. We will discuss the case of multi-body particle decays, for which the spin states of daughter particles will be referred relatively the mother $A$ particle states, defined in a given $A$ rest frame reference system $S_A$, by specifying a sequence of Lorentz transformations.

Following Ref.~\cite{Leader2011}, we consider active Lorentz transformations on the physical system: particle momenta are boosted and rotated, while the reference coordinate system $(T,X,Y,Z)$ stays unchanged and the spatial directions specify the spin coordinate systems (\ie the set of spin operators) of daughter particles in their rest frame. Particle momenta therefore define the spin coordinate system assigned to a given particle via the sequence of Lorentz transformations applied. To keep track of the Lorentz transformations applied, the following notation is used: a reference system $S'$ defined in terms of another $S$ system by applying the Lorentz transformation $\Lambda$ to the momenta expressed in the $S$ system is indicated as $S'=\Lambda S$.

The choice of the spin coordinate system for a particle $B$ moving with momentum $\bm{p}^A(B)$ in $S_A$ is ambiguous, since $B$ particle rest frames obtained by Lorentz transformations differing by rotations around $\bm{p}^A(B)$ represent the particle spin with different spin states\footnote{In this article, we consider the definition of the spin coordinate system for a particle at rest from that of a reference system in relative motion, while in Ref.~\cite{Leader2011} it is the spin coordinate system of the moving reference frame to be defined from that of the particle at rest. The two approaches are equivalent, but we adopt the first being more suitable for the case of particle decays, in which the initial spin coordinate system is given for the mother, rather than the daughter particles.}. Note that the time in the particle $B$ frame is uniquely defined by the boost connecting different rest frames, its definition considered implicitly in the following.
There are two main choices of spin coordinate systems in literature~\cite{Leader2011}:
\begin{itemize}
\item The canonical system $S^C_B$ is the spin coordinate system obtained from $S_A$ by doing a Lorentz boost $L\left[-\bm{p}^A(B)\right]$,
\begin{equation}
S^C_B = L\left[-\bm{p}^A(B)\right] S_A.
\label{eq:canoncial_state_definition}
\end{equation}
The canonical coordinate system is physically determined by the $A$ momentum in the $B$ particle rest frame $\bm{p}^B(A)$ requiring its direction to be opposite to that of $\bm{p}^A(B)$ in the $A$ coordinate system,
\begin{equation}
\frac{\bm{p}_{x_i}^B(A)}{p^B(A)} = -\frac{\bm{p}_{X_i}^A(B)}{p^A(B)}.
\end{equation}
\item The helicity system $S^H_B$ consists in choosing the particle spin quantisation axis to be opposite to the $\bm{p}^B(A)$ direction. The helicity system for $\bm{p}^A(B)$ having polar and azimuthal angles $\theta$ and $\phi$ in the $S_A$ system\footnote{The function $\mathrm{atan2}(y,x) \in [-\pi,\pi]$ computes the signed angle between the $x$ axis and the vector having components $(x,y)$.},
\begin{align}
\cos\theta &= \left(p_Z^A(B)/|p^A(B)|\right), \nonumber\\
\phi &= \mathrm{atan2}\left(p_Y^A(B), p_X^A(B) \right),
\label{eq:helicity_angles}
\end{align}
is obtained by applying a $R(-\psi,-\theta,-\phi)$ Euler rotation, defined in Appendix~\ref{sec:euler_rotations}, aligning $\bm{p}^A(B)$ with the $Z$ axis, followed by a boost $L(-p^A(B) \bm{Z})$,
\begin{equation}
S^H_B = L(-p^A(B) \bm{Z})R(-\psi,-\theta,-\phi) S_A.
\label{eq:helicity_state_definition}
\end{equation}
With this definition, the particle spin states are described in terms of the helicity states $\Ket{\bm{p}^A(B),s,\lambda}$ introduced in Sect.~\ref{sec:phase_definition}.
The angle $\psi$, associated to a rotation around $\bm{p}^A(B)$, determines the choice of the orthogonal spin coordinate axes. The orthogonal axes can be chosen arbitrarily, but once defined must be consistently specified to avoid introducing phase differences, as shown in Sect.~\ref{sec:phase_definition}.
Their definition it is better visualised as a passive rotation $R(\phi,\theta,\psi)$ applied on the $(X,Y,Z)$ coordinate system. If there are particles other than $A$ and $B$ involved in the process, their momenta will provide a physical definition of the orthogonal axes, since they will change under the $\psi$ angle rotation.

Note that any choice of the $\psi$ angle is valid, even if the two used in the literature are $\psi=0$ and $\psi=-\phi$.
In the rest of the paper, the simplest choice $\psi=0$ will be employed for the definition of helicity states.
\end{itemize}

A simple relation holds between helicity and canonical systems: the boost $L\left[-\bm{p}^A(B)\right]$ can be decomposed as
\begin{equation}
L\left[-\bm{p}^A(B)\right] = R(\phi,\theta,0)L(-p^A(B) \bm{Z})R(0,-\theta,-\phi),
\label{eq:boost_decomposition}
\end{equation}
so that
\begin{equation}
S^C_B = R(\phi,\theta,0) S^H_B.
\label{eq:canonical_helicity_relation}
\end{equation}
The relation between canonical $\Ket{\bm{p}^A(B),s,m}$ and helicity $\Ket{\bm{p}^A(B),s,\lambda}$ spin states, which are both defined in particle B rest frames, is given by the Wigner $D$-matrix (introduced in Appendix~\ref{sec:euler_rotations}) representing the active rotation from the helicity to the canonical system,
\begin{equation}
R(\phi,\theta,0)\Ket{\bm{p}^A(B),s,\lambda} = \sum_{m} D^{s}_{m,\lambda}(\phi,\theta,0) \Ket{\bm{p}^A(B),s,m}.
\label{eq:canonical_helicity_relation_states}
\end{equation}


Note that both canonical and helicity states are defined relatively to the given reference system $S_A$: canonical and helicity states defined starting from different reference systems differ by the so-called Wigner and Wick rotations, respectively~\cite{Leader2011}. When considering a chain of spin reference systems linked by canonical or helicity transformations, as in the case of multi-body particle decays, the relation among spin states defined in the first and the last systems depends on the whole sequence of transformations. This fact has physical consequences, the most famous being Thomas precession~\cite{Thomas:1926dy,*Thomas:1927yu}; in Sect.~\ref{sec:helicity_amplitudes} we show how it is important to take into account the whole sequence of transformations for a correct definition of decay amplitudes.

\section{Euler rotations and their representation on spin states}
\label{sec:euler_rotations}
Here, we introduce Euler rotations and their representation on spin states, in non-relativistic quantum mechanics, following the conventions of Ref.~\cite{Richman}. We consider an active rotation on a physical system described by a reference coordinate system $(x,y,z)$. The rotation is described by introducing a new coordinate system $(x',y',z')$ which is rotated with respect to the reference one together with the physical system. A generic rotation of a physical system can be described by means of Euler rotations, which, in the $z$-$y$-$z$ convention for the rotation axes, are defined as
\begin{align}
R(\alpha,\beta,\gamma) = R_{z}(\alpha) R_{y}(\beta) R_{z}(\gamma) = e^{-i\alpha \hat{S}_{z}} e^{-i\beta \hat{S}_{y}} e^{-i\gamma \hat{S}_{z}},
\label{eq:euler_rotation}
\end{align}
in which rotations are expressed with respect to the original axes $(x,y,z)$. The three Euler angles $\alpha,\beta,\gamma$ can be computed from the unit vectors describing the rotated system in the terms of the original coordinates as
\begin{align}
\alpha &= \mathrm{atan2} \left( z'_y, z'_x \right) \in [-\pi,\pi], \nonumber\\
\beta &= \arccos\left( z'_z \right) \in [0,\pi], \nonumber\\
\gamma &= \mathrm{atan2} \left( y'_z, - x'_z \right) \in [-\pi,\pi].
\label{eq:euler_angles}
\end{align}

The action of an Euler rotation on spin states $\ket{s,m}$ associated to the spin coordinate system $(x,y,z)$ is
\begin{equation}
\hat{R}(\alpha,\beta,\gamma)\ket{s,m} = \sum_{m'=-s}^{s} D^s_{m',m}(\alpha,\beta,\gamma)\ket{s,m'},
\label{eq:euler_rotation_spin_states}
\end{equation}
in which the Wigner $D$-matrices $D^s_{m',m}(\alpha,\beta,\gamma)$ are
\begin{equation}
D^{s}_{m',m}(\alpha,\beta,\gamma) = \braket{s,m'|R(\alpha,\beta,\gamma)|s,m}.
\label{eq:Wigner_D_matrix_definition}
\end{equation}
The Wigner $D$-matrices of index $s$ are spin $s$ representations of the $SU(2)$ group.
Following Eq.~\eqref{eq:euler_rotation}, the Wigner $D$-matrices can be factorised as
\begin{align}
D^{s}_{m',m}(\alpha,\beta,\gamma) = \braket{s,m'|e^{-i\alpha \hat{S}_{z}} e^{-i\beta \hat{S}_{y}} e^{-i\gamma \hat{S}_{z}}|s,m}= e^{-im\alpha}d^s_{m',m}(\beta)e^{-im'\gamma},
\label{eq:Wigner_d_matrix_definition}
\end{align}
in which the Wigner $d$-matrix elements are real combinations of trigonometric functions of $\beta$, their analytical expression reported in Ref.~\cite{Richman}. Wigner $D$-matrices have many properties following from those of the rotation group; for the purpose of this article we report
\begin{align}
D^{s}_{m',m}(\alpha,\beta,\gamma) &= D^{s}_{m,m'}(\gamma,-\beta,\alpha),\nonumber\\
D^{*s}_{m',m}(\alpha,\beta,\gamma) &= D^{s}_{m',m}(-\alpha,\beta,-\gamma).
\label{eq:D_matrix_properties}
\end{align}

The inverse rotation from the final coordinate system $(x',y',z')$ to the initial one $(x,y,z)$ follows from Eq.~\eqref{eq:euler_rotation},
\begin{align}
R^{-1}(\alpha,\beta,\gamma) &= R_{z}^{-1}(\gamma) R_{y}^{-1}(\beta) R_{z}^{-1}(\alpha)
= e^{i\gamma \hat{S}_{z}} e^{i\beta \hat{S}_{y}} e^{i\alpha \hat{S}_{z}}\nonumber\\
&=R(-\gamma,-\beta,-\alpha),
\end{align}
and the Wigner $D$-matrix representation on $\ket{s,m}$ states is $D^{s}_{m',m}(-\gamma,-\beta,-\alpha)$. This definition of inverse Euler rotation allows to undo step-by-step the three rotations composing the Euler rotation. The usual inverse Euler rotation, given in terms of positive $\beta$ angle, invert the rotation following a different path, which can introduce additional $2\pi$ rotations leading to inequivalent trajectories in the $SU(2)$ group representing spin. This issue is relevant for fermion states, as mentioned in Sect.~\ref{sec:phase_definition}.

As a practical example, let's consider the inverse of a rotation of angle $\theta$ around the $y$ axis, $R(0,\theta,0)$, which can be chosen to be $R(0,-\theta,0)$ or $R(\pi,\theta,\pi)$. The Wigner $D$-matrix representing $\hat{R}(0,-\theta,0)$ on a spin 1/2 state is
\begin{equation}
D^{1/2}_{m',m}(0,-\theta,0) = d^{1/2}_{m',m}(-\theta) = \left(
\begin{array}{ccc}
\cos\frac{\theta}{2} & & \sin\frac{\theta}{2}\\[2ex]
-\sin\frac{\theta}{2} & & \cos\frac{\theta}{2}\\
\end{array}
\right),
\end{equation}
while that representing $\hat{R}(\pi,\theta,\pi)$ is
\begin{align}
D^{1/2}_{m',m}(\pi,\theta,\pi) &= e^{-i\pi(m+m')} d^{1/2}_{m',m}(-\theta)\nonumber\\
&= \left(
\begin{array}{ccc}
-\cos\frac{\theta}{2} & & -\sin\frac{\theta}{2}\\[2ex]
\sin\frac{\theta}{2} & & -\cos\frac{\theta}{2}\\
\end{array}
\right) \nonumber\\
&= -D^{1/2}_{m',m}(0,-\theta,0),
\end{align}
which has opposite sign with respect to the first matrix.

\section{Polarised differential decay rate}
\label{sec:polarised_rate}
The generic spin state of a statistical ensemble of particles is defined by the associated density operator $\hat{\rho}$~\cite{Leader2011}: given an ensemble of spin states $\Ket{\psi}_i$ occurring with probability $p_i$, the density operator is
\begin{equation}
\hat{\rho} = \sum_i p_i \Ket{\psi}_i\Bra{\psi}_i,
\label{eq:density_matrix}
\end{equation}
so that the expectation value of any operator $\hat{X}$ can be expressed as
\begin{equation}
\langle\hat{X}\rangle = \sum_i p_i \Braket{\psi|\hat{X}|\psi}_i = \mathrm{Tr}\left[\hat{\rho}\hat{X}\right].
\end{equation}
From the definition Eq.~\eqref{eq:density_matrix} follows that the density operator is hermitian with unit trace.

The decay rate of multi-body decays $A\to \lbrace i=1,..,n \rbrace$ for definite spin eigenstates is the squared modulus of the transition amplitude between the $A$ particle initial state $\Ket{s_A,m_A}$ and the final particle product state $\Ket{\lbrace s_i\rbrace,\lbrace m_i\rbrace} = \otimes_i \Ket{s_i,m_i}$,
\begin{align}
p_{m_A,\lbrace m_i\rbrace}(\Omega) &= |\braket{s_A,m_A|\hat{T}|\lbrace s_i\rbrace,\lbrace m_i\rbrace}|^2 \nonumber\\
&= |\mathcal{A}_{m_A,\lbrace m_i\rbrace}(\Omega)|^2.
\label{eq:decay_rate_pure_states}
\end{align}
Generic polarisation states are described by introducing the density operators for the initial particle state $\hat{\rho}^A$ and the final particle states $\hat{\rho}^{\lbrace i\rbrace}$, which are included in the decay rate Eq.~\eqref{eq:decay_rate_pure_states} by inserting suitable identity resolutions, obtaining
\begin{align}
p(\hat{\rho}^A,\hat{\rho}^{\lbrace i\rbrace};\Omega) &= \mathrm{tr} \left[ \hat{\rho}^A \hat{T} \hat{\rho}^{\lbrace i\rbrace} \hat{T}^{\dagger} \right]\nonumber\\
&= \sum_{m_A,m'_A} \sum_{\lbrace m_i\rbrace,\lbrace m'_i\rbrace} \hat{\rho}^A_{m_A,m'_A} \hat{\rho}^{\lbrace i\rbrace}_{\lbrace m_i\rbrace,\lbrace m'_i\rbrace} \mathcal{A}_{m_A,\lbrace m_i\rbrace}(\Omega) \mathcal{A}^*_{m'_A,\lbrace m'_i\rbrace}(\Omega).
\label{eq:decay_rate_mixed_states}
\end{align}

\section{\Lcpkpi amplitude model}
\label{sec:Lcpkpi_model}
The \Lcpkpi amplitude model is built closely following the three-body helicity amplitudes described in Sect.~\ref{sec:helicity_amplitudes}, with the identifications $A\leftrightarrow \Lc$, $1\leftrightarrow p$, $2\leftrightarrow K^-$, $3\leftrightarrow \pi^+$, $R\leftrightarrow \Lz^*$, $S\leftrightarrow \Deltares^{*++}$, $U\leftrightarrow K^*$. In this appendix we report the definition of the helicity amplitudes for the specific \Lcpkpi decay case.

Starting from the decay chain $\Lc\to pK^*(\to K^-\pi^+)$, the weak decay $\Lc\to pK^*$ is described by
\begin{equation}
\mathcal{A}^{\Lc\to pK^*}_{m_{\Lc},\lambda_p,\bar{\lambda}_{K^*}} = \mathcal{H}^{\Lc\to pK^* }_{\lambda_p,\bar{\lambda}_{K^*}} D^{*1/2}_{m_{\Lc},\lambda_p+\bar{\lambda}_{K^*}}(\phi_p,\theta_p,0),
\end{equation}
in which proton and $K^*$ helicities $\lambda_p$ and $\bar{\lambda}_{K^*}$ are defined in the proton helicity frame reached from the \Lc baryon polarisation frame. For spin zero $K^*$ resonances the angular momentum conservation relations Eq.~\eqref{eq:allowed_helicity_couplings} allow two complex couplings corresponding to $m_p=\pm 1/2$, for higher spin resonances four couplings are allowed, corresponding to $\lbrace m_p=1/2$; $\bar{\lambda}_{K^*}=0,-1\rbrace$ and $\lbrace m_p=-1/2$; $\bar{\lambda}_{K^*}=0,1 \rbrace$. The couplings are independent of each other because of parity violation in weak decays. The strong decay $K^*\to K^-\pi^+$ contribution is 
\begin{equation}
\mathcal{A}^{K^*\to K^-\pi^+}_{\bar{\lambda}_{K^*}} = \mathcal{H}^{K^*\to K^-\pi^+}_{0,0} D^{*J_{K^*}}_{\bar{\lambda}_{K^*},0}(\bar{\phi}_K,\bar{\theta}_K,0)\mathcal{R}(\mqkpi),
\end{equation}
in which $\bar{\phi}_{K}$ and $\bar{\theta}_{K}$ are the kaon azimuthal and polar angles in the $K^*$ opposite-helicity frame (obtained boosting to the $K^*$ rest frame after the rotation $R(\phi_p,\theta_p,0)$). The function $\mathcal{R}(\mqkpi)$ describes the non-negligible mass width of the intermediate state. The coupling $\mathcal{H}^{K^*\to K^-\pi^+}_{0,0}$ can not be determined independently of $\mathcal{H}^{\Lc\to K^* p}_{m_p,\bar{\lambda}_{K^*}}$ couplings, thus it is set equal to 1 and absorbed into the latter.

Considering the decay chain $\Lc\to\Lz^*(\to pK^-)\pi^+$, the weak decay $\Lc\to\Lz^*\pi^+$ is described by
\begin{equation}
\mathcal{A}^{\Lc\to\Lz^*\pi^+}_{m_{\Lc},\lambda_{\Lz^*}} = \mathcal{H}^{\Lc\to\Lz^*\pi^+}_{\lambda_{\Lz^*},0} D^{*1/2}_{m_{\Lc},\lambda_{\Lz^*}}(\phi_{\Lz^*},\theta_{\Lz^*},0),
\end{equation}
with $\lambda_{\Lz^*}$ defined in the $\Lz^*$ helicity system reached from the \Lc reference frame. The angular momentum conservation relations Eq.~\eqref{eq:allowed_helicity_couplings} allow two helicity couplings, $\lambda_{\Lz^*}=\pm 1/2$, to fit for each resonance whatever $J_{\Lz^*}$ is. The strong decay $\Lz^*\to pK^-$ is described by
\begin{equation}
\mathcal{A}^{\Lz^*\to pK^-}_{\lambda_{\Lz^*},\lambda^{\Lz^*}_p} = \mathcal{H}^{\Lz^*\to pK^-}_{\lambda^{\Lz^*}_p,0} D^{*J_{\Lz^*}}_{\lambda_{\Lz^*},\lambda^{\Lz^*}_p}(\phi^{\Lz^*}_p,\theta^{\Lz^*}_p,0)\mathcal{R}(\mqpk),
\end{equation}
with $\lambda^{\Lz^*}_p$ the proton helicity in the system reached from the $\Lz^*$ helicity frame. Parity conservation in strong decays relates the two helicity couplings corresponding to $\lambda^{\Lz^*}_p=\pm 1/2$ by
\begin{equation}
\mathcal{H}^{\Lz^*\to pK^-}_{-\lambda^{\Lz^*}_p,0} = -P_{\Lz^*} (-1)^{J_{\Lz^*}-1/2} \mathcal{H}^{\Lz^*\to pK^-}_{\lambda^{\Lz^*}_p,0},
\end{equation}
in which $P_{\Lz^*}$ is the parity of the $\Lz^*$ resonance and the proton and kaon parities $P_p=1$, $P_K=-1$ have been inserted. They are absorbed into $\mathcal{H}^{\Lc\to\Lz^*\pi^+}_{\lambda_{\Lz^*},0}$ couplings by setting
\begin{equation}
\mathcal{H}^{\Lz^*\to pK^-}_{+1/2,0}=1, \hspace{24pt}
\mathcal{H}^{\Lz^*\to pK^-}_{-1/2,0}=-P_{\Lz^*} (-1)^{J_{\Lz^*}-1/2}.
\end{equation}

Considering the third decay chain $\Lc\to\Deltares^{++*}(\to p\pi^+)K^-$, the weak decay $\Lc\to\Deltares^{++*}K^-$ is described by
\begin{equation}
\mathcal{A}^{\Lc\to\Deltares^*K^-}_{m_{\Lc},\lambda_{\Deltares^*}} = \mathcal{H}^{\Lc\to\Deltares^*K^-}_{\lambda_{\Deltares^*},0} D^{*1/2}_{m_{\Lc},\lambda_{\Deltares^*}}(\phi_{\Deltares^*},\theta_{\Deltares^*},0),
\end{equation}
with $\lambda_{\Deltares^*}$ defined in the $\Deltares^*$ helicity system reached from the \Lc frame. The strong decay $\Deltares^{++*}\to p\pi^+$ amplitude is written as
\begin{equation}
\mathcal{A}^{\Deltares^*\to p\pi^+}_{\lambda_{\Deltares^*},\lambda^{\Deltares^*}_p} = \mathcal{H}^{\Deltares^*\to p\pi^+}_{\lambda^{\Deltares^*}_p,0} D^{*J_{\Deltares^*}}_{\lambda_{\Deltares^*},\lambda^{\Deltares^*}_p}(\phi^{\Deltares^*}_p,\theta^{\Deltares^*}_p,0)\mathcal{R}(\mqppi),
\end{equation}
with $\lambda^{\Deltares^*}_p$ defined in the proton helicity system reached from the $\Deltares^*$ helicity frame. Helicity couplings are defined as for the $\Lz^*$ decay chain.

The decay amplitudes for each decay chain are obtained as in Eqs.~\eqref{eq:helicity_amplitude_3body_R_state} and~\eqref{eq:helicity_amplitude_3body_S_state}, to which the rotations needed to match the proton spin state definitions must be applied, as in Eqs.~\eqref{eq:helicity_amplitude_3body_R_state_rotated} and~\eqref{eq:helicity_amplitude_3body_S_state_rotated}.

For the numerical study presented in Sect.~\ref{sec:effects_phase} we consider a \Lcpkpi amplitude model consisting of three resonances, one per decay channel, with invariant mass dependence described by relativistic Breit-Wigner functions, characterised by mass and width parameters. We consider the following spin-parity $J^P$ assignments: $K^*(1^-)$, $\Lz^*(1/2^-)$ and $\Deltares^*(1/2^-)$. The values of the complex couplings and Breit-Wigner mass and width parameters employed in the study are reported in Table~\ref{tab:toy_fit_pol}:  they are chosen in order to produce significant interference effects.

\begin{table}
\centering
\begin{tabular}{lcccc}
\toprule
Parameter & Value\\
\midrule
$K^*$\\
$\mathcal{H}_{1/2,0}$ & $1$\\
$\mathcal{H}_{1/2,-1}$ & $0.5 + 0.5i$\\
$\mathcal{H}_{-1/2,1}$ & $i$ &\\
$\mathcal{H}_{-1/2,0}$ & $- 0.5 - 0.5i$\\
$m(\gev)$ & 0.9\\
$\Gamma(\gev)$ & 0.2\\
\midrule
$\Lz^*$\\
$\mathcal{H}_{-1/2,0}$ & $i$\\
$\mathcal{H}_{1/2,0}$ & $0.8 - 0.4i$\\
$m(\gev)$ & 1.6\\
$\Gamma(\gev)$ & 0.2\\
\midrule
$\Deltares^*$\\
$\mathcal{H}_{-1/2,0}$ & $0.6 - 0.4i$\\
$\mathcal{H}_{1/2,0}$ & $0.1i$\\
$m(\gev)$ & 1.4\\
$\Gamma(\gev)$ & 0.2\\
\bottomrule
\end{tabular}
\caption{Helicity couplings and Breit-Wigner parameter values employed in the numerical study of Sect.~\ref{sec:effects_phase}. \label{tab:toy_fit_pol}}
\end{table}


\setboolean{inbibliography}{true}
\bibliographystyle{LHCb}
\bibliography{biblio}   


\end{document}